\documentclass[12pt,a4paper]{article}
\usepackage[latin1]{inputenc}
\usepackage{amsmath}
\usepackage{amsfonts}
\usepackage{amssymb}
\author{Jussi I. Tyhtila}
\title{Analytical solution for 3-dimensional, incompressible Navier-Stokes equations with a suitable external force field}

\begin{document}
\maketitle
jussi.i.lindgren@helsinki.fi \\
+358443700704
Malminkatu 36 C 61 00100 Helsinki
\begin{abstract}
This paper provides primarily an analytical \textit{ad hoc} -solution for 3-dimensional, incompressible Navier-Stokes equations with a suitable external force field. The solution turns out to be smooth and integrable, as there is a gaussian decay of amplitudes present. 
\end{abstract}

\section{Introduction}
This article presents and ad hoc -solution to full Navier-stokes system in incompressible case. We begin in a didactic way by presenting a rather trivial solution to Navier-Stokes system, then we proceed with a more complex situation. The trivial solution is spatially periodic and smooth, the more complex solution turns out to be integrable on the whole space, also it is found to be a smooth one. The solution is obtained considering a divergence free vector field with gaussian decay of amplitudes, and finally adding a suitable force field. The mathematical treatment is rather elementary, mainly vector calculus. One can refer e.g. \cite{VC}. On the details of Navier-Stokes, one can see e.g \cite{FD}.
\newpage
\section{Navier-Stokes equations}
The incompressible Navier-stokes equations are in vector form
\begin{equation}
\frac{\partial \vec{u}}{\partial t}+\nabla p=\nu\triangle \vec{u}-\vec{u}\cdot \nabla \vec{u} +\vec{f}
\end{equation}
\begin{equation}
\nabla \cdot \vec{u}=0
\end{equation}
where $\vec{u}=(u_x(x,y,z,t),u_y(x,y,z,t),u_z(x,y,z,t))$ is a vector (velocity) field $\vec{u}:\mathbb{R}^3\times [0,\infty)\longrightarrow \mathbb{R}^3$, $\vec{f}=(f_x(x,y,z,t),f_y(x,y,z,t),f_z(x,y,z,t))$ is a vector (force) field $\vec{f}:\mathbb{R}^3\times [0,\infty)\longrightarrow \mathbb{R}^3$ and $p(x,y,z,t)$ is a scalar (pressure) field $p:\mathbb{R}^3\times [0,\infty)\longrightarrow \mathbb{R}$. $\nu>0$ (viscosity) is a constant. \\
In terms of Cartesian coordinates, the equations can be stated equivalently as
\begin{equation}
\frac{\partial u_x}{\partial t}+\frac{\partial p}{\partial x}=\nu(\frac{\partial^2 u_x}{\partial x^2}+\frac{\partial^2 u_x}{\partial y^2}+\frac{\partial^2 u_x}{\partial z^2})-u_x\frac{\partial u_x}{\partial x}-u_y\frac{\partial u_x}{\partial y}-u_z\frac{\partial u_x}{\partial z}+f_x
\end{equation}
\begin{equation}
\frac{\partial u_y}{\partial t}+\frac{\partial p}{\partial y}=\nu(\frac{\partial^2 u_y}{\partial x^2}+\frac{\partial^2 u_y}{\partial y^2}+\frac{\partial^2 u_y}{\partial z^2})-u_x\frac{\partial u_y}{\partial x}-u_y\frac{\partial u_y}{\partial y}-u_z\frac{\partial u_y}{\partial z} +f_y
\end{equation}
\begin{equation}
\frac{\partial u_z}{\partial t}+\frac{\partial p}{\partial z}=\nu(\frac{\partial^2 u_z}{\partial x^2}+\frac{\partial^2 u_z}{\partial y^2}+\frac{\partial^2 u_z}{\partial z^2})-u_x\frac{\partial u_z}{\partial x}-u_y\frac{\partial u_z}{\partial y}-u_z\frac{\partial u_z}{\partial z} +f_z
\end{equation}
\begin{equation}
\frac{\partial u_x}{\partial x}+\frac{\partial u_y}{\partial y}+\frac{\partial u_z}{\partial z}=0
\end{equation}

\section{Introductory preliminaries;solution of Navier-Stokes with no external force field}
Consider the functions
\begin{equation}
u_x(x,y,z,t)=e^{at+ i(\alpha x+\beta y+\gamma z)}
\end{equation}
\begin{equation}
u_y(x,y,z,t)=e^{at+ i(\alpha x+\beta y+\gamma z)}
\end{equation}
\begin{equation}
u_z(x,y,z,t)=e^{at+ i(\alpha x+\beta y+\gamma z)}
\end{equation}
with $x,y,z\in \mathbb{R}$, $t\geq 0$ and $a,\alpha,\beta,\gamma \in \mathbb{R}$. \\
We thus have $u_x=u_y=u_z=u$, equivalently $\vec{u}=(u(x,y,z,t),u(x,y,z,t),u(x,y,z,t))$. We shall adopt this notation.

Let us calculate the partial derivatives. We will have the following functional equations for $u_x$

$$
\frac{\partial u_x}{\partial x}=i\alpha u_x \; ,\; \frac{\partial u_x}{\partial y}=i\beta u_x \; ,\; \frac{\partial u_x}{\partial z}=i\gamma u_x
$$
$$
\frac{\partial^2u_x}{\partial x^2}=-\alpha^2 u_x \; ,\; \frac{\partial^2u_x}{\partial y^2}=-\beta^2 u_x \; ,\; \frac{\partial^2u_x}{\partial z^2}=-\gamma^2 u_x
$$

Similarly for $u_y$
$$
\frac{\partial u_y}{\partial x}=i\alpha u_y \; ,\; \frac{\partial u_y}{\partial y}=i\beta u_y \; ,\; \frac{\partial u_y}{\partial z}=i\gamma u_y
$$
$$
\frac{\partial^2u_y}{\partial x^2}=-\alpha^2 u_y \; ,\; \frac{\partial^2u_y}{\partial y^2}=-\beta^2 u_y \; ,\; \frac{\partial^2u_y}{\partial z^2}=-\gamma^2 u_y
$$

And for $u_z$
$$
\frac{\partial u_z}{\partial x}=i\alpha u_z \; ,\; \frac{\partial u_z}{\partial y}=i\beta u_z \; ,\; \frac{\partial u_z}{\partial z}=i\gamma u_z
$$
$$
\frac{\partial^2u_z}{\partial x^2}=-\alpha^2 u_z \; ,\; \frac{\partial^2u_z}{\partial y^2}=-\beta^2 u_z \; ,\; \frac{\partial^2u_z}{\partial z^2}=-\gamma^2 u_z
$$
From these we find that
\begin{equation}
\nabla \cdot \vec{u}=\frac{\partial u_x}{\partial x}+\frac{\partial u_y}{\partial y}+\frac{\partial u_z}{\partial z}=i\alpha u_x+i\beta u_y+i\gamma u_z=0
\end{equation}
As $u_x=u_y=u_z=u$, we will have
\begin{equation}
iu(\alpha+\beta +\gamma)=0
\end{equation}
From which we conclude
\begin{equation}
\alpha +\beta+ \gamma=0
\end{equation}
We shall call it the \textit{incompressibility condition}. \\
Let us substitute the partial derivatives into Navier-Stokes equations.We will use the notation $u=u_x=u_y=u_z$. We will obtain
\begin{equation}
\frac{\partial u}{\partial t}+\frac{\partial p}{\partial x}=-\nu(\alpha^2 u+\beta^2 u+\gamma^2 u)-i\alpha u^2-i\beta u^2-i\gamma u^2
\end{equation}
\begin{equation}
\frac{\partial u}{\partial t}+\frac{\partial p}{\partial y}=-\nu(\alpha^2 u+\beta^2 u+\gamma^2 u)-i\alpha u^2-i\beta u^2-i\gamma u^2
\end{equation}
\begin{equation}
\frac{\partial u}{\partial t}+\frac{\partial p}{\partial z}=-\nu(\alpha^2 u+\beta^2 u+\gamma^2 u)-i\alpha u^2-i\beta u^2-i\gamma u^2
\end{equation}
This is called The Fundamental Functional Equation.

Which equal
\begin{equation}
\frac{\partial u}{\partial t}+\frac{\partial p}{\partial x}=-\nu(\alpha^2 +\beta^2 +\gamma^2 )u-iu^{2}(\alpha + \beta +\gamma)
\end{equation}
\begin{equation}
\frac{\partial u}{\partial t}+\frac{\partial p}{\partial y}=-\nu(\alpha^2 +\beta^2 +\gamma^2 )u-iu^{2}(\alpha + \beta +\gamma)
\end{equation}
\begin{equation}
\frac{\partial u}{\partial t}+\frac{\partial p}{\partial z}=-\nu(\alpha^2 +\beta^2 +\gamma^2 )u-iu^{2}(\alpha + \beta +\gamma)
\end{equation}

Now, as from the incompressibility condition $\alpha + \beta +\gamma =0$, we will have just
\begin{equation}
\frac{\partial u}{\partial t}+\frac{\partial p}{\partial x}=-\nu(\alpha^2 +\beta^2 +\gamma^2 )u
\end{equation}
\begin{equation}
\frac{\partial u}{\partial t}+\frac{\partial p}{\partial y}=-\nu(\alpha^2 +\beta^2 +\gamma^2 )u
\end{equation}
\begin{equation}
\frac{\partial u}{\partial t}+\frac{\partial p}{\partial z}=-\nu(\alpha^2 +\beta^2 +\gamma^2 )u
\end{equation}

This is the fundamental functional equation. Let us turn to calculate the time-derivative. As $u(x,y,z,t)=e^{at+ i(\alpha x+\beta y+\gamma z)}$, we will have
\begin{equation}
\frac{\partial u}{\partial t}=au
\end{equation}
From which we will have the parametric functional equation
\begin{equation}
au+\frac{\partial p}{\partial x}=-\nu(\alpha^2 +\beta^2 +\gamma^2 )u
\end{equation}
\begin{equation}
au+\frac{\partial p}{\partial y}=-\nu(\alpha^2 +\beta^2 +\gamma^2 )u
\end{equation}
\begin{equation}
au+\frac{\partial p}{\partial z}=-\nu(\alpha^2 +\beta^2 +\gamma^2 )u
\end{equation}

This holds, if 
\begin{equation}
a=-\nu(\alpha^2 +\beta^2 +\gamma^2 ) \; \; \; and \; \; \; \frac{\partial p}{\partial x}=\frac{\partial p}{\partial y}=\frac{\partial p}{\partial z}=0\Longrightarrow p(x,y,z,t)=c
\end{equation}
$c\in\mathbb{R}$. \\
This gives the final solution
\begin{equation}
u(x,y,z,t)=e^{-\nu(\alpha^2 +\beta^2 +\gamma^2 )t+ i(\alpha x+\beta y+\gamma z)}
\end{equation}
with the restriction $\alpha+\beta+\gamma=0$.
Dropping the imaginary component, we will have
\begin{equation}
u(x,y,z,t)=e^{-\nu(\alpha^2 +\beta^2 +\gamma^2 )t}cos(\alpha x+\beta y+\gamma z)
\end{equation}
with the restriction $\alpha + \beta +\gamma =0$ and
\begin{equation}
p(x,y,z,t)=c, \; \; c\in \mathbb{R}
\end{equation}

\section{Analytical solution for Navier-Stokes equations}

\subsection{The 'grundfunctions'}
We shall begin with the arbitrary choice of 'grundfunctions', as we shall call them. They are completely ad hoc in their nature.
\begin{equation}
u_x(x,y,z,t)= \alpha (Biy-y)(Ciz-z)e^{at-\frac{1}{2}(x^2+y^2+z^2)+\frac{i}{2}(Ax^2+By^2+Cz^2)}
\end{equation}
\begin{equation}
u_y(x,y,z,t)=\beta (Aix-x)(Ciz-z)e^{at-\frac{1}{2}(x^2+y^2+z^2)+\frac{i}{2}(Ax^2+By^2+Cz^2)}
\end{equation}
\begin{equation}
u_z(x,y,z,t)=\gamma (Aix-x)(Biy-y)e^{at-\frac{1}{2}(x^2+y^2+z^2)+\frac{i}{2}(Ax^2+By^2+Cz^2)}
\end{equation}
with $a,A,B,C,\alpha ,\beta , \gamma \in \mathbb{C}$ and $i$ is the imaginary unit. \\

For things to come, we calculate straightforwardly
\begin{equation}
u_xu_y=\alpha \beta (Aix-x)(Biy-y)(Ciz-z)^2e^{2at-(x^2+y^2+z^2)+i(Ax^2+By^2+Cz^2)}
\end{equation}
\begin{equation}
u_xu_z=\alpha \gamma (Aix-x)(Biy-y)^2(Ciz-z)e^{2at-(x^2+y^2+z^2)+i(Ax^2+By^2+Cz^2)}
\end{equation}
\begin{equation}
u_yu_z=\beta \gamma (Aix-x)^2(Biy-y)(Ciz-z)e^{2at-(x^2+y^2+z^2)+i(Ax^2+By^2+Cz^2)}
\end{equation}

We note immediately that the divergence of such vector field of grundfunctions vanishes, as
\begin{equation}
\nabla \cdot \vec{u}=(Aix-x)u_x+(Biy-y)u_y+(Ciz-z)u_z=0 \forall x,y,z,t
\end{equation}
if and only if
\begin{equation}
\alpha +\beta +\gamma=0
\end{equation}
This is the incompressibility condition. \\
\subsection{Derivatives of grundfunctions}
Let us calculate the partial derivatives. We will have the following functional equations for $u_x$

$$
\frac{\partial u_x}{\partial x}=(Aix-x)u_x
$$
$$
\frac{\partial u_x}{\partial y}=(Biy-y)u_x+\frac{u_x}{y}
$$
$$
\frac{\partial u_x}{\partial z}=(Ciz-z)u_x+\frac{u_x}{z}
$$

The second partial derivatives
$$
\frac{\partial^2 u_x}{\partial x^2}=(Aix-x)^2u_x+(Ai-1)u_x
$$
$$
\frac{\partial^2 u_x}{\partial y^2}=(Biy-y)^2u_x+3(Bi-1)u_x
$$
$$
\frac{\partial^2 u_x}{\partial z^2}=(Ciz-z)^2u_x+3(Ci-1)u_x 
$$
The cross-terms
$$
u_x\frac{\partial u_x}{\partial x}=(Aix-x)u_{x}^{2}
$$
$$
u_y\frac{\partial u_x}{\partial y}=(Biy-y)u_xu_y+\frac{u_xu_y}{y}
$$
$$
u_z\frac{\partial u_x}{\partial z}=(Ciz-z)u_xu_z+\frac{u_xu_z}{z}
$$
So sum of cross-terms will be
\begin{equation}
u_x((Aix-x)u_{x}+(Biy-y)u_y+\frac{u_y}{y}+(Ciz-z)u_z+\frac{u_z}{z})=\frac{u_xu_y}{y}+\frac{u_xu_z}{z}
\end{equation}

And for $u_y$

$$
\frac{\partial u_y}{\partial x}=(Aix-x)u_y+\frac{u_y}{x}
$$
$$
\frac{\partial u_y}{\partial y}=(Biy-y)u_y
$$
$$
\frac{\partial u_y}{\partial z}=(Ciz-z)u_y+\frac{u_y}{z}
$$

The second partial derivatives
$$
\frac{\partial^2 u_y}{\partial x^2}=(Aix-x)^2u_y+3(Ai-1)u_y
$$
$$
\frac{\partial^2 u_y}{\partial y^2}=(Biy-y)^2u_y+(Bi-1)u_y
$$
$$
\frac{\partial^2 u_y}{\partial z^2}=(Ciz-z)^2u_y+3(Ci-1)u_y
$$
The cross terms
$$
u_x\frac{\partial u_y}{\partial x}=u_x((Aix-x)u_y+\frac{u_y}{x})
$$
$$
u_y\frac{\partial u_y}{\partial y}=(Biy-y)u_{y}^{2}
$$
$$
u_z\frac{\partial u_y}{\partial z}=u_z((Ciz-z)u_y+\frac{u_y}{z})
$$
So sum of cross-terms will be
\begin{equation}
u_y((Aix-x)u_{x}+(Biy-y)u_y+\frac{u_x}{x}+(Ciz-z)u_z+\frac{u_z}{z})=\frac{u_yu_x}{x}+\frac{u_yu_z}{z}
\end{equation}

And for $u_z$

$$
\frac{\partial u_z}{\partial x}=(Aix-x)u_z+\frac{u_z}{x}
$$
$$
\frac{\partial u_z}{\partial y}=(Biy-y)u_z+\frac{u_z}{y}
$$
$$
\frac{\partial u_z}{\partial z}=(Ciz-z)u_z
$$

The second partial derivatives
$$
\frac{\partial^2 u_z}{\partial x^2}=(Aix-x)^2u_z+3(Ai-1)u_z
$$
$$
\frac{\partial^2 u_z}{\partial y^2}=(Biy-y)^2u_z+3(Bi-1)u_z
$$
$$
\frac{\partial^2 u_z}{\partial z^2}=(Ciz-z)^2u_x+(Ci-1)u_x
$$
The cross-terms
$$
u_x\frac{\partial u_z}{\partial x}=u_x((Aix-x)u_z+\frac{u_z}{x})
$$
$$
u_y\frac{\partial u_z}{\partial y}=u_y((Biy-y)u_z+\frac{u_z}{y})
$$
$$
u_z\frac{\partial u_z}{\partial z}=(Ciz-z)u_{z}^{2}
$$
So sum of cross-terms will be
\begin{equation}
u_z((Aix-x)u_{x}+(Biy-y)u_y+\frac{u_x}{x}+(Ciz-z)u_z+\frac{u_y}{y})=\frac{u_zu_x}{x}+\frac{u_zu_y}{y}
\end{equation}
\subsection{Navier-Stokes equations with the choice of grundfunctions}

The right side of the Navier-Stokes -system will become
\begin{equation}
\nu((Aix-x)^2+(Ai-1)+(Biy-y)^2+3(Bi-1)+(Ciz-z)^2+3(Ci-1) )u_x-(\frac{u_xu_y}{y}+\frac{u_xu_z}{z}) +f_x
\end{equation}
\begin{equation}
\nu((Aix-x)^2+3(Ai-1)+(Biy-y)^2+(Bi-1)+(Ciz-z)^2+3(Ci-1))u_y-(\frac{u_yu_x}{x}+\frac{u_yu_z}{z}) +f_y
\end{equation}
\begin{equation}
\nu((Aix-x)^2+3(Ai-1)+(Biy-y)^2+3(Bi-1)+(Ciz-z)^2+(Ci-1))u_z-(\frac{u_zu_x}{x}+\frac{u_zu_y}{y}) + f_z
\end{equation}

The time derivatives will be obviously
\begin{equation}
\frac{\partial u_x}{\partial t}=au_x
\end{equation}
\begin{equation}
\frac{\partial u_y}{\partial t}=au_y
\end{equation}
\begin{equation}
\frac{\partial u_z}{\partial t}=au_z
\end{equation}
Therefore we naturally choose,
\begin{equation}
a=\nu((A+B+C)i-3)
\end{equation}

As we want $a$ to be a constant. Therefore the grundfunctions will become
\begin{equation}
u_x(x,y,z,t)= \alpha (Biy-y)(Ciz-z)e^{\nu((A+B+C)i-3)t-\frac{1}{2}(x^2+y^2+z^2)+\frac{i}{2}(Ax^2+By^2+Cz^2)}
\end{equation}
\begin{equation}
u_y(x,y,z,t)=\beta (Aix-x)(Ciz-z)e^{\nu((A+B+C)i-3)t-\frac{1}{2}(x^2+y^2+z^2)+\frac{i}{2}(Ax^2+By^2+Cz^2)}
\end{equation}
\begin{equation}
u_z(x,y,z,t)=\gamma (Aix-x)(Biy-y)e^{\nu((A+B+C)i-3)t-\frac{1}{2}(x^2+y^2+z^2)+\frac{i}{2}(Ax^2+By^2+Cz^2)}
\end{equation}
One can see that the the field is rotating with respect to time due to the factor $(A+B+C)i$. \\

Therefore the remaining pressure gradient will have components
\begin{equation}
\frac{\partial p}{\partial x}=\nu((Aix-x)^2+(Biy-y)^2+2(Bi-1)+(Ciz-z)^2+2(Ci-1))u_x-(\frac{u_xu_y}{y}+\frac{u_xu_z}{z}) +f_x
\end{equation}
\begin{equation}
\frac{\partial p}{\partial y}=\nu((Aix-x)^2+2(Ai-1)+(Biy-y)^2+(Ciz-z)^2+2(Ci-1))u_y-(\frac{u_yu_x}{x}+\frac{u_yu_z}{z}) +f_y
\end{equation}
\begin{equation}
\frac{\partial p}{\partial z}=\nu((Aix-x)^2+2(Ai-1)+(Biy-y)^2+2(Bi-1)+(Ciz-z)^2)u_z-(\frac{u_zu_x}{x}+\frac{u_zu_y}{y}) + f_z
\end{equation}
We do not know in general, whether such pressure field exists, instead, we construct an explicit solution.

\subsection{Explicit solution with a nontrivial external force field}
We will show an explicit solution with the aid of an external force field. We consider an external force field
\begin{equation}
\vec{f}=(f_x,f_y,f_z)
\end{equation}
We assume it is smooth and integrable. 
For simplicity we will take $A=B=C=1$, that is, the frequencies are all of unity. Now consider the pressure gradient. \\
\begin{equation}
\frac{\partial p}{\partial x}=\nu((ix-x)^2u_x+(iy-y)^2u_x+2(i-1)u_x+(iz-z)^2u_x+2(i-1)u_x)-(\frac{u_xu_y}{y}+\frac{u_xu_z}{z})+f_x
\end{equation}
\begin{equation}
\frac{\partial p}{\partial y}=\nu((ix-x)^2u_y+2(i-1)u_y+(iy-y)^2u_y+(iz-z)^2u_y+2(i-1)u_y)-(\frac{u_yu_x}{x}+\frac{u_yu_z}{z})+f_y
\end{equation}
\begin{equation}
\frac{\partial p}{\partial z}=\nu((ix-x)^2u_z+2(i-1)u_z+(iy-y)^2u_z+2(i-1)u_z+(iz-z)^2u_z)-(\frac{u_zu_x}{x}+\frac{u_zu_y}{y})+f_z
\end{equation}

Let us multiply the first equation by $(xi-x)$, the second by $(yi-y)$ and the third by $(zi-z)$, and sum the three equations with each other. Using the property of zero-divergence, we will get just
\begin{equation}
-x(i-1)(\frac{u_xu_y}{y}+\frac{u_xu_z}{z}-f_x)-y(i-1)(\frac{u_yu_x}{x}+\frac{u_yu_z}{z}-f_y)-z(i-1)(\frac{u_zu_x}{x}+\frac{u_zu_y}{y}-f_z)=
\end{equation}
\begin{equation}
x(i-1)\frac{\partial p}{\partial x}+y(i-1)\frac{\partial p}{\partial y}+z(i-1)\frac{\partial p}{\partial z}
\end{equation}
Let us choose backwards so that
\begin{equation}
\frac{\partial p}{\partial x}=-(\frac{u_xu_y}{y}+\frac{u_xu_z}{z})+f_x
\end{equation}
\begin{equation}
\frac{\partial p}{\partial y}=-(\frac{u_yu_x}{x}+\frac{u_yu_z}{z})+f_y
\end{equation}
\begin{equation}
\frac{\partial p}{\partial z}=-(\frac{u_zu_x}{x}+\frac{u_zu_y}{y})+f_z
\end{equation}

As we have
\begin{equation}
\frac{u_xu_y}{y}=-4\alpha \beta xz^2e^{\nu(6i-6)t-(x^2+y^2+z^2)+i(x^2+y^2+z^2)}
\end{equation}

\begin{equation}
\frac{u_xu_z}{z}=-4\alpha \gamma xy^2e^{\nu(6i-6)t-(x^2+y^2+z^2)+i(x^2+y^2+z^2)}
\end{equation}

\begin{equation}
\frac{u_zu_y}{y}=-4\beta \gamma zx^2e^{\nu(6i-6)t-(x^2+y^2+z^2)+i(x^2+y^2+z^2)}
\end{equation}

\begin{equation}
\frac{u_yu_x}{x}=-4\beta \alpha yz^2e^{\nu(6i-6)t-(x^2+y^2+z^2)+i(x^2+y^2+z^2)}
\end{equation}

\begin{equation}
\frac{u_zu_x}{x}=-4\alpha \gamma zy^2e^{\nu(6i-6)t-(x^2+y^2+z^2)+i(x+y+z)}
\end{equation}

\begin{equation}
\frac{u_yu_z}{z}=-4\beta \gamma yx^2e^{\nu(6i-6)t-(x^2+y^2+z^2)+i(x^2+y^2+z^2)}
\end{equation}

With $g=e^{\nu(6i-6)t-(x^2+y^2+z^2)+i(x^2+y^2+z^2)}$

\begin{equation}
\frac{\partial p}{\partial x}=4\alpha( \beta z^2+\gamma y^2)xg +f_x
\end{equation}
\begin{equation}
\frac{\partial p}{\partial y}=4\beta(\alpha  z^2+\gamma x^2)yg +f_y
\end{equation}
\begin{equation}
\frac{\partial p}{\partial z}=4\gamma(\alpha  y^2+\beta  x^2)zg+f_z
\end{equation}

Now suppose we have an external force field $\vec{f}$, such that
\begin{equation}
f_x=-4\alpha( \beta z^2+\gamma y^2)xg+4xg
\end{equation}
\begin{equation}
f_y=-4\beta(\alpha  z^2+\gamma x^2)yg+4yg
\end{equation}
\begin{equation}
f_z=-4\gamma(\alpha  y^2+\beta  x^2)zg+4zg
\end{equation}

Then the pressure gradient will be just
\begin{equation}
\frac{\partial p}{\partial x}=4xg
\end{equation}
\begin{equation}
\frac{\partial p}{\partial y}=4yg
\end{equation}
\begin{equation}
\frac{\partial p}{\partial z}=4zg
\end{equation}
This can be integrated easily to obtain
\begin{equation}
p(x,y,z,t)=-(1+i)e^{\nu(6i-6)t-(x^2+y^2+z^2)+i(x^2+y^2+z^2)}+ constant
\end{equation}

\subsection{Conclusion}
We have provided an analytical solution to the full Navier-Stokes equations that is smooth and integrable with the aid of an external force field. Now one can proceed to extend this solution to more complex situations. What kind of frequency combinations are possible? Are there nontrivial solutions without an external force field?

\end{document}